\documentclass[aps,prl,superscriptaddress,amsfonts,amsmath,amssymb,showpacs,floatfix,reprint]{revtex4-1}

\usepackage{bm}
\usepackage{graphicx}
\usepackage{amsmath}
\usepackage{amstext}
\usepackage{amssymb}
\usepackage{amsfonts}
\usepackage{amsbsy}
\usepackage{verbatim}
\usepackage[usenames,dvipsnames,svgnames,table]{xcolor}
\usepackage{multirow}
\usepackage{gensymb}
\usepackage{textcomp}
\usepackage{enumitem}
\usepackage{comment}
\usepackage{fancyhdr}
\usepackage[version=3]{mhchem}
\usepackage[colorlinks=true, urlcolor=Blue, linkcolor=Blue, citecolor=Blue, pdftex]{hyperref}

\newcommand{\lu}{Lu$_2$Mo$_2$O$_5$N$_2$}

\begin{document}

\title{Signatures of a gearwheel quantum spin liquid in a spin-$\frac{1}{2}$ pyrochlore molybdate Heisenberg antiferromagnet}

\author{Yasir Iqbal}
\email{yiqbal@physics.iitm.ac.in}
\affiliation{Department of Physics, Indian Institute of Technology Madras, Chennai, 600036, India}
\author{Tobias M\"uller}
\affiliation{Institute for Theoretical Physics and Astrophysics, Julius-Maximilian's
University of W\"urzburg, Am Hubland, D-97074 W\"urzburg, Germany}
\author{Kira Riedl}
\affiliation{Institut f\"ur Theoretische Physik, Goethe-Universit{\"a}t Frankfurt,
Max-von-Laue-Stra{\ss}e 1, D-60438 Frankfurt am Main, Germany}
\author{Johannes Reuther}
\affiliation{Dahlem Center for Complex Quantum Systems and Fachbereich Physik, Freie Universit{\"a}t Berlin, D-14195 Berlin, Germany}
\affiliation{Helmholtz-Zentrum Berlin f{\"u}r Materialien und Energie, D-14109 Berlin, Germany}
\author{Stephan Rachel}
\affiliation{School of Physics, The University of Melbourne, Parkville, VIC 3010, Australia}
\affiliation{Institut f\"ur Theoretische Physik, Technische Universit\"at Dresden, D-01062 Dresden, Germany}
\author{Roser Valent\'i}
\affiliation{Institut f\"ur Theoretische Physik, Goethe-Universit{\"a}t Frankfurt,
Max-von-Laue-Stra{\ss}e 1, D-60438 Frankfurt am Main, Germany}
\author{Michel J. P. Gingras}
\affiliation{Perimeter Institute for Theoretical Physics, Waterloo, Ontario, Canada N2L 5G7}
\affiliation{Department of Physics and Astronomy, University of Waterloo, Waterloo, Ontario, Canada N2L 3G1}
\affiliation{Canadian Institute for Advanced Research, 180 Dundas Street West, Toronto, Ontario, Canada M5G 1Z8}
\author{Ronny Thomale}
\affiliation{Institute for Theoretical Physics and Astrophysics, Julius-Maximilian's
University of W\"urzburg, Am Hubland, D-97074 W\"urzburg, Germany}
\author{Harald O. Jeschke}
\affiliation{Research Institute for Interdisciplinary Science, Okayama University, 
3-1-1 Tsushima-naka, Kita-ku, Okayama 700-8530, Japan}

\date{\today}

\begin{abstract}
 We theoretically investigate the low-temperature phase of the recently synthesized {\lu} material, an extraordinarily rare realization of a $S=1/2$ three-dimensional pyrochlore Heisenberg antiferromagnet in which Mo$^{5+}$ are the $S=1/2$ magnetic species. Despite a Curie-Weiss temperature ($\Theta_{\rm CW}$) of $-121(1)$ K, experiments have found no signature of magnetic ordering \emph{or} spin freezing down to $T^*\approx0.5$ K. Using density functional theory, we find that the compound is well described by a Heisenberg model with exchange parameters up to third nearest neighbors. The analysis of this model via the pseudofermion functional renormalization group method reveals paramagnetic behavior down to a temperature of at least $T=|\Theta_{\rm CW}|/100$, in agreement with the experimental findings hinting at a possible three-dimensional quantum spin liquid. The spin susceptibility profile in reciprocal space shows momentum-dependent features forming a ``gearwheel'' pattern, characterizing what may be viewed as a molten version of a chiral noncoplanar incommensurate spiral order under the action of quantum fluctuations. Our calculated reciprocal space susceptibility maps provide benchmarks for future neutron scattering experiments on single crystals of {\lu}.
\end{abstract}

\maketitle

{\it Introduction}. A quantum spin liquid (QSL) is an exotic strongly correlated paramagnetic quantum state of matter~\cite{Pomeranchuk-1941,Anderson-1973,Fazekas-1974} that lacks conventional long-range magnetic order 
down to absolute zero temperature and is characterized by nontrivial spin entanglement and low-energy fractionalized spin excitations~\cite{Balents-2010,Savary-2017,Zhou-2017}.
One of the ideal settings to explore QSL physics is provided by systems in which the magnetic moments reside on either a two- or three-dimensional network of corner-shared (CS) triangles or tetrahedra and interact with an isotropic nearest-neighbor antiferromagnetic Heisenberg exchange Hamiltonian. The promise of such systems stems, in part, from their low propensity to order even at the classical level~\cite{Villain-1979,Chalker-1992,Moessner-1998b}. Materials with magnetic species described by an (effective) $S=1/2$ operator are expected to display the most extreme quantum behaviors, as suggested by numerous theoretical and numerical works spanning over $25$ years~\cite{Harris-1991,Canals-1998,Erez-2003,Huang-2016,Yan-2011,Iqbal-2013,Norman-2016,He-2017,Liao-2017}, and are manifestly of significant interest.

While one might legitimately expect that single-ion anisotropy and exchange anisotropy would much 
undermine the likeliness of a QSL, the proposals that QSL states may be
realized in systems described by effective $S=1/2$ degrees of freedom, but
with strongly anisotropic bilinear spin-spin couplings originating from large
spin-orbit interactions, are exciting developments in the field. These include
``Kitaev''
materials~\cite{Jackeli-2009,Reuther-2011d,Winter-2016,Hermanns-2017,
Trebst-2017,Winter-2017}
based on Ir$^{4+}$ or Ru$^{3+}$, and ``quantum spin ice''
(QSI)~\cite{Molavian-2007,Onoda-2010,Ross-2011,Gingras-2014} pyrochlore oxide
materials with trivalent rare-earth ions. 

In the above Heisenberg antiferromagnets, Kitaev and QSI systems, one has at
 hand a reference (idealized) Hamiltonian ${\cal H}_0$ as the model presumed
 to host a QSL state.  The general mindset in the field has been to
 consider materials whose true Hamiltonian, 
 ${\cal H}={\cal H}_0 + {\cal H}^\prime$, may not 
be ``too far'' from  ${\cal H}_0$
 in terms of all material-relevant perturbations ${\cal H}^\prime$.
From a material perspective, the search and discovery of QSL phases 
thus require some luck so that
${\cal H}^\prime$ is sufficiently weak that long-range order is evaded. The experimental investigation of such potential QSL materials requires the synthesis of single crystals which, albeit being at times a daunting challenge, is a necessary one as it allows to expose the nontrivial momentum dispersion of low-energy excitations characterizing QSL states~\cite{Han-2012,Hao-2013,Punk-2013}.

A prime candidate for a QSL phase in two dimensions is the herbertsmithite kagome material
where long-range exchange beyond nearest neighbor as well as  the
Dzyaloshinsky-Moriya (DM) interaction
might be subcritical to drive this compound to 
a magnetic long-range ordered state~\cite{Mendels-2007,Han-2012,Zorko-2017}.
Illustrating further the subcritical role of 
further interactions, one may note the 
kapellasite kagome compound~\cite{Fak-2012,Kermarrec-2014}, which is altogether described at ``zeroth order'' by a complex spin Hamiltonian with numerous competing interactions beyond nearest neighbor, landing it in a parameter space island where a QSL may be realized
~\cite{Bernu-2013,Jeschke-2013,Iqbal-2015}.
Nevertheless, the number of candidates for QSL behavior in two dimensions
is small and the situation for three-dimensional materials is even more disconcerting. The pyrochlore lattice of CS tetrahedra, occurring in pyrochlore oxides and spinel magnetic materials, is an attractive architecture to search for QSLs~\cite{Harris-1991,Canals-1998,Erez-2003,Nussinov-2007,Banerjee-2008,Shannon-2010,Normand-2014,Normand-2016,Huang-2016}.
Unfortunately, most materials in these two families either develop long-range
magnetic order or display a spin-glass-like freezing at low temperature, hence
averting a QSL state. Similarly, Na$_4$Ir$_3$O$_8$~\cite{Okamoto-2007}, an
antiferromagnetic spin-$\frac{1}{2}$ material with a three-dimensional
hyperkagome lattice of CS triangles, also exhibits a spin freezing below about
7 K~\cite{Shockley-2015}. The MgTi$_2$O$_4$ spinel has $S=1/2$ Ti$^{3+}$
moments, but structurally distorts at low temperature~\cite{Isobe-2002}.
Finally, most
Kitaev materials so far identified display long-range order and the
 behaviors of the best QSI candidates remain far from being well rationalized~\cite{Trebst-2017}.

 One may thus offer an executive summary of the experimental
situation, especially for three-dimensional materials:
 In all cases, the perturbations
${\cal H}^\prime$ are above a critical value and preempt the formation of a
QSL.  At this juncture, a convergence of opportunities, from the point of view
of (i) potential QSL material candidates and (ii) an ability to model its
${\cal H}$ and expose its QSL nature with  state-of-the-art numerical methods,
is required to encourage the significant efforts in the synthesis of pertinent
single crystals of three-dimensional QSL candidates. In this context, we propose in this Rapid Communication that {\lu} is a candidate much deserving such effort and subsequent investigation.
  
Lu$_2$Mo$_2$O$_5$N$_2$ is a pyrochlore Heisenberg antiferromagnet with Mo$^{5+}$ $S=1/2$ moments that fail to develop long-range order \emph{or} spin freezing down to $T^*\approx0.5$ K, despite a Curie-Weiss temperature of $\Theta_{\rm CW}=-121(1)$ K~\cite{Clark-2014}. Notwithstanding the appeal of its $S=1/2$ ${\cal H}_0$ Heisenberg antiferromagnetic nature, we characterize in this work the leading perturbation ${\cal H}^\prime$ of {\lu} in the hope of identifying a material with an innocuous $\mathcal{H}'$ such that it does not induce long-range magnetic order.

\begin{figure}
\includegraphics[width=\columnwidth]{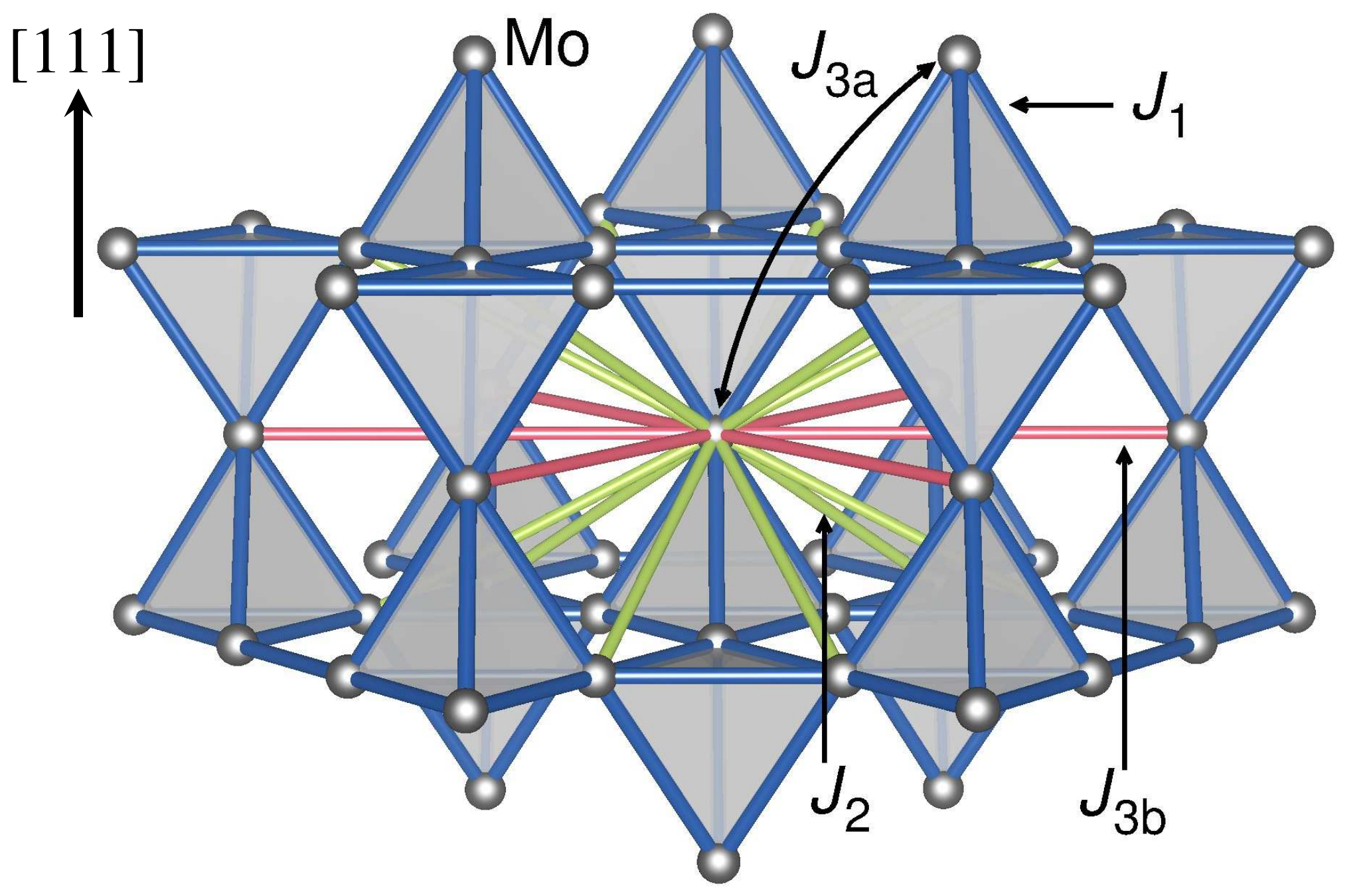}
\caption{Leading exchange paths in the pyrochlore lattice of {\lu}. Only the Mo$^{5+}$ ions are shown.}\label{fig:structure}
\end{figure}

While the nonmagnetic random site O/N disorder might certainly be worth considering at a later stage, in this Rapid Communication, as a first step in fleshing out the leading physics at play in {\lu}, we model this material as an effective homogeneous $S=1/2$ pyrochlore magnet. We apply a combination of (i) density functional theory (DFT) determination of the Hamiltonian parameters where the random O/N occupation is modelled using the virtual crystal approximation~\cite{Singh-2007}, (ii) a $S=1/2$ pseudofermion functional renormalization group (PFFRG) study of the resulting Heisenberg Hamiltonian, and (iii) an analysis of the multiple-$\mathbf{k}$ spiral order that is realized for a classical version of the spin model derived from DFT. We establish the nature of the perturbation $\mathcal{H}'$ and find it to be \emph{meek} at inducing long-range order\textemdash likely the one key factor for the failure of this material to freeze or order down to $\vert T^*/\Theta_{\rm CW}\vert {\ll} 1$. It is shown that the long-range (third-nearest-neighbor) exchange coupling, in particular, $J_{3\rm a}$ [see Fig.~\ref{fig:structure} and Eq.~(\ref{eqn:Hamil})], is crucial for defining a minimal material-relevant model Hamiltonian for {\lu}, as found for chromium spinels~\cite{Yaresko-2008,Tymoshenko-2017}. For the model of Eq.~(\ref{eqn:Hamil}) below, the PFFRG shows an absence of magnetic order down to temperatures $T^{*}\approxeq|\Theta_{\rm CW}|/100$, in agreement with experiment. A classical analysis~\cite{Nakamura-2007,Ioki-2007,Conlon-2010,Okubo-2011,Lapa-2012} of this model identifies a noncoplanar triple-$\mathbf{k}$ incommensurate spiral order as the parent classical state, whose melting by quantum fluctuations, would give a suitable phenomenological frame to describe the observed quantum spin liquid, possibly of chiral nature, and its $\mathbf{k}$-dependent spin susceptibility fingerprint.

\begin{figure}
\includegraphics[width=\columnwidth]{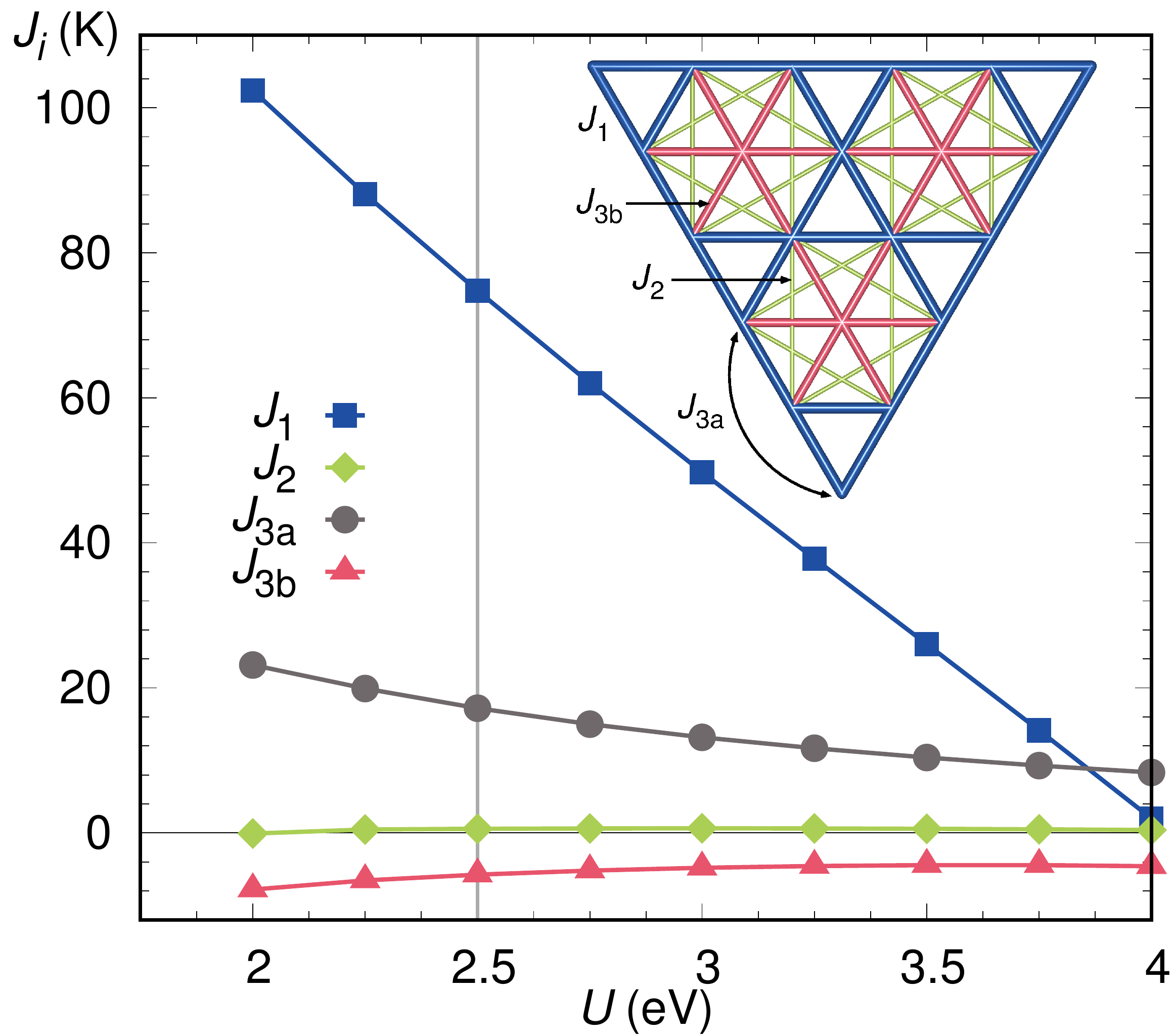}
\caption{Calculated exchange couplings for {\lu} as given in Table~\ref{tab:DFT}~\cite{supp}. Positive (negative) values correspond to antiferromagnetic (ferromagnetic) couplings. A generalized gradient approximation (GGA+$U$) functional with $J_{\rm H}=0.6$~eV was used. A vertical line marks the set of couplings which corresponds to a Curie-Weiss temperature of $\Theta_{\rm CW}=-125(4)$~K, in good agreement with the experimental value of $-121(1)$ K. The inset shows a detail of the magnetic lattice with the first four exchange paths between Mo$^{5+}$ ions, as viewed from the $[111]$ direction (see Fig.~\ref{fig:structure}).}\label{fig:couplings}
\end{figure}

\begin{figure*}
\includegraphics[width=2.0\columnwidth]{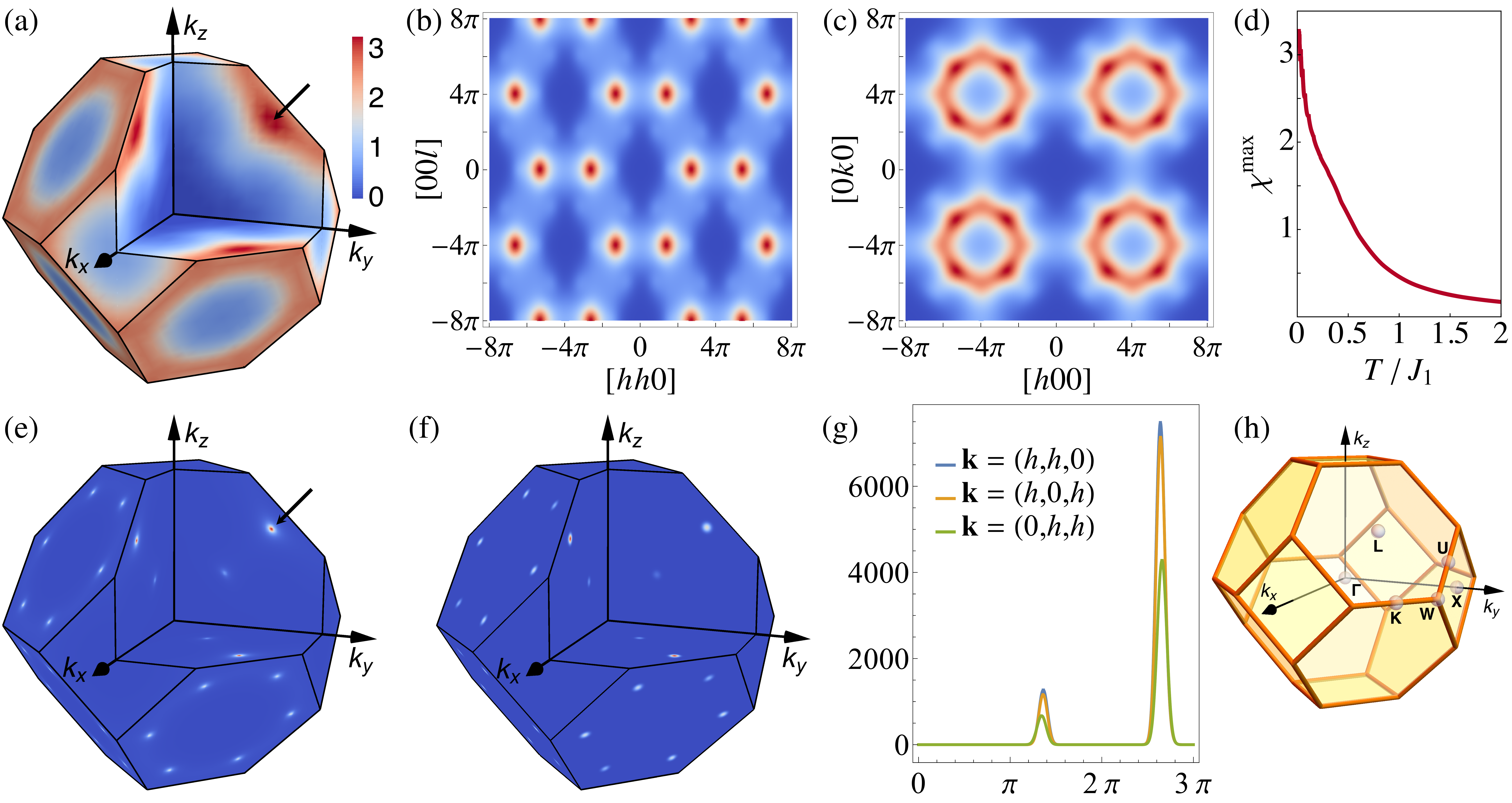}
\caption{\label{fig:susc}
First row: $S=1/2$ PFFRG simulation of the model Hamiltonian [Eq.~(\ref{eqn:Hamil})] for \lu. The magnetic susceptibility (in units of $1/J_{1}$) is shown at $T=|\Theta_{\rm CW}|/100$ in the (a) full Brillouin zone, (b) $[hhl]$ plane, and (c) $[hk0]$ plane. (d) The RG flow of the dominant susceptibility. Second row: (e)\textendash(g) Susceptibility profiles of the corresponding classical spin model obtained from (e) $S{\to}\infty$ limit of PFFRG, (f) iterative energy minimization, also shown along (g) selected cuts. (h) Brillouin zone of the pyrochlore lattice, a truncated octahedron, illustrating the high-symmetry points.}
\end{figure*}

{\it Results}. The minimal model for {\lu} extracted from our DFT calculations~\cite{Jeschke-2011,Jeschke-2013} is given by a four-parameter isotropic $S=1/2$ Heisenberg model,
\begin{eqnarray}\label{eqn:Hamil}
\hat{{\cal H}} &=& J_1 \sum_{{\langle i,j \rangle_{1}}} \mathbf{\hat{S}}_{i} \cdot \mathbf{\hat{S}}_{j}
+J_2 \sum_{{\langle i,j \rangle_{2}}} \mathbf{\hat{S}}_{i} \cdot \mathbf{\hat{S}}_{j}\nonumber \\
&+&J_{3\rm a} \sum_{{\langle i,j \rangle_{3{\rm a}}}} \mathbf{\hat{S}}_{i} \cdot \mathbf{\hat{S}}_{j}
+J_{3\rm b} \sum_{{\langle i,j \rangle_{3{\rm b}}}} \mathbf{\hat{S}}_{i} \cdot \mathbf{\hat{S}}_{j},
\end{eqnarray}
where ${\mathbf{\hat{S}}}_{i}$ is a quantum spin-$\frac{1}{2}$ operator at pyrochlore lattice site $i$. The indices ${\langle i,j \rangle_{1(2)}}$ denote sums over nearest-neighbor (second-nearest-neighbor) pairs of sites. There are two inequivalent third-nearest-neighbor sites, the ${\langle i,j \rangle_{3\rm a}}$ (connecting two Mo$^{5+}$ sites with a nearest-neighbor Mo$^{5+}$ ion in between) and ${\langle i,j \rangle_{3\rm b}}$ (across an empty hexagon in one of the three interpenetrating kagome lattices of the pyrochlore structure) (see Fig.~\ref{fig:structure}). 
We find that $J_1,J_2,J_{3{\rm a}}>0$ are antiferromagnetic while $J_{3{\rm b}}<0$ is ferromagnetic (see Fig.~\ref{fig:structure}). The set of exchange couplings corresponding to $U=2.5$ eV (see Table~\ref{tab:DFT}~\cite{supp} and Fig.~\ref{fig:couplings}) give an estimate of the Curie-Weiss temperature $\Theta_{\rm CW}=-125(4)$ K corresponding to the experimentally determined value of $\Theta_{\rm CW}=-121(1)$ K. The couplings are found to be $(J_{2},J_{3\rm a},J_{3\rm b})=(0.008,0.23,-0.078)$ in units of $J_1$, with $J_{1}=1$.

The PFFRG~\cite{Reuther-2010,Reuther-2011a,Reuther-2011b,Reuther-2011d,Metzner-2012,Iqbal-2016b,Finn-2016,Hering-2017,Baez-2016} calculations (see Ref.~\cite{supp}) for the model Hamiltonian [Eq.~(\ref{eqn:Hamil})] for {\lu} were performed on a cluster of 2315 correlated sites with the longest spin-spin correlator being $\sim$$11.5$ nearest-neighbor lattice spacings, which ensures an adequate $\mathbf{k}$-space resolution. The $\mathbf{k}$-space resolved spin susceptibility profile evaluated at the lowest temperature ($T=|\Theta_{\rm CW}|/100=1.21$ K) is shown in Fig.~\ref{fig:susc}(a). At a temperature which is two orders of magnitude smaller compared to $\Theta_{\rm {CW}}$, the diffused spectral weight along the edges of the Brillouin zone (with a slight enhancement at the W points) reflects the high degree of frustration in {\lu}. Interestingly, analogous features in the spectral weight distribution around the boundary are also shared by the highly frustrated spin-$\frac{1}{2}$ kagome Heisenberg antiferromagnet~\cite{Depenbrock-2012,Iqbal-2013,Suttner-2014}. Away from the boundaries, one observes soft maxima [marked by an arrow in Fig.~\ref{fig:susc}(a)] at an incommensurate wave vector ${\mathbf k}_{\rm QSL}=2\pi(1.296,1.296,0)$ (and symmetry-related points). The $\mathbf{k}$-dependent features of the susceptibility are best visualized in the $[hhl]$ plane, i.e., $k_{x}=k_{y}$ plane [Fig.~\ref{fig:susc}(b)]. Therein, we observe that the spectral weight at the pinch points [$(h,h,l)=(0,0,4\pi)$ in Fig.~\ref{fig:susc}(b)] is both substantially suppressed and smeared and, instead,  redistributes to form hexagonal clusters~\cite{Fennell-2008}, similar to what is observed in ZnCr$_{2}$O$_{4}$~\cite{Lee-2002}. This behavior is a consequence of the nonzero third-nearest-neighbor couplings $J_{3\rm a}$ and $J_{3\rm b}$ in Eq.~(\ref{eqn:Hamil}), as has been argued in Ref.~\cite{Conlon-2010} on the basis of a classical analysis. In the $[hk0]$ plane, i.e., $k_{z}=0$ plane [Fig.~\ref{fig:susc}(c)], the characteristic spin susceptibility profile resembles a pattern of ``gearwheels'' and, following Ref.~\cite{Okumura-2010}, we dub the spin liquid accordingly. The RG flow of the susceptibility tracked at the dominant wave vector ${\mathbf k}_{\rm QSL}$ is shown in Fig.~\ref{fig:susc}(d)~\footnote{The value of $\mathbf{k}_{\rm QSL}$ changes only minimally with temperature. The RG flows tracked at these different $\mathbf{k}_{\rm QSL}$ vectors all show paramagnetic behavior.}, wherein the observed  oscillations at small temperature arise due to frequency discretization. Its monotonic increase as $T{\to} 0$ without any indication of a divergence points to the absence of a magnetic phase transition, in agreement with experiment~\cite{Clark-2014}. We reach similar conclusions for exchange couplings corresponding to different values of $U$ in the range $2~{\rm eV}\leqslant U\leqslant 3.25~{\rm eV}$ given in Table~\ref{tab:DFT}~\cite{supp}.

In order to identify the classical long-range magnetic order associated with Eq.~(\ref{eqn:Hamil}), we use both the PFFRG method, and an iterative energy minimization of the classical Hamiltonian~\cite{Lapa-2012}. In the $S\to\infty$ limit, the PFFRG flow equations permit an exact analytic solution in the thermodynamic limit and the approach is equivalent to the Luttinger-Tisza method~\cite{Baez-2016}. The resulting ground states on non-Bravais lattices are approximate, since only the global constraint $\sum_{i}|\mathbf{S}_{i}^{2}|=S^{2}N$, where $N$ is the total number of lattice sites, is enforced~\cite{Nussinov-2004,Kimchi-2014}. We find that under the RG flow, the two-particle vertex for the magnetic ordering (MO) wave vector, ${\mathbf k_{\rm MO}=2\pi(1.305,1.305,0)}$ [marked by an arrow in Fig.~\ref{fig:susc}(e)] (and symmetry-related points), diverges at a N\'eel temperature of $T_{N}/J_{1} \approx 0.625$, denoting the onset of an incommensurate magnetic order. The susceptibility profile evaluated at this ordering temperature is shown in Fig.~\ref{fig:susc}(e). One observes that the susceptibility profile of the $S=1/2$ model [Fig.~\ref{fig:susc}(a)] may be viewed as a diffuse version of the one for the classical model [Fig.~\ref{fig:susc}(e)]. Under the action of quantum fluctuations, the subdominant Bragg peaks on the hexagonal faces in Fig.~\ref{fig:susc}(e) become diffuse to form a uniform ring in Fig.~\ref{fig:susc}(a), while the dominant Bragg peaks at $\mathbf{k}_{\rm MO}$ smear out to form a gearwheel pattern, albeit leaving behind fingerprints [marked by an arrow in Fig.~\ref{fig:susc}(a)]. The whitish ``teeth'' of the gearwheels seen in Fig.~\ref{fig:susc}(c) can, likewise, be accounted for. 

To obtain the exact classical ground state and, in addition, allow for possible lattice symmetry breaking, we perform an iterative classical energy minimization enforcing the constraint $|\mathbf{S}_{i}|^{2}=S^{2}$ at each site $i$~\cite{Kimchi-2014}. This yields a magnetic state that is a noncoplanar triple-$\mathbf{k}$ structure composed of a superposition of three different spirals, each governed by an incommensurate wave vector $\mathbf{k}$. Moreover, we find that although the total spin per tetrahedron is not zero, the deviation is not energetically significant, being only a few percent of $J_{1}$. This implies an approximate equivalence between the antiferromagnetic $J_{3\rm a}$ and ferromagnetic $J_{2}$ couplings~\cite{Chern-2008,Lapa-2012}, and accounts for the similarities of the orders found here with those of the $J_{1}$-$J_{2}$ Heisenberg model~\cite{Nakamura-2007,Ioki-2007,Okubo-2011}. The corresponding susceptibility profile is shown in Fig.~\ref{fig:susc}(f), with the dominant Bragg peaks located at $\mathbf{k}^{'}_{\rm MO}=2\pi(1.312,1.312,0)$ (in good agreement with $\mathbf{k}_{\rm MO}$) (and symmetry-related points). The finite-size effects due to periodic boundary conditions cause Bragg peak splitting, and the results in Figs.~\ref{fig:susc}(f) and ~\ref{fig:susc}(g) are shown after performing a Gaussian smoothing over the split peaks. It is important to note that the height of the Bragg peaks in the $k_{x}$-$k_{y}$ and $k_{x}$-$k_{z}$ planes are slightly different, but are roughly twice the height of the peak in the $k_{y}$-$k_{z}$ plane [see Fig.~\ref{fig:susc}(g)]. One may wonder whether the breaking of the cubic symmetry in the classical order could carry over to the $S=1/2$ case and give rise to a nematic QSL~\cite{Iqbal-2016a,Iqbal-2016c}. 

Interestingly, the spin configuration of our classical magnetic order is chiral, namely, that the effect of a time-reversal operation $\mathbf{S}{\to}\mathbf{-S}$ cannot be undone by a global SO(3) spin rotation. This is precisely the defining characteristic of a {\emph chiral} spin state~\cite{Messio-2011} which, accordingly, exhibits a nonvanishing scalar spin chirality $\sim\mathbf{S}_{i}\cdot(\mathbf{S}_{j}\times\mathbf{S}_{k}$). Indeed, we find that on every tetrahedron, any set of three spins gives a nonzero scalar spin chirality. The prospect of this chiral symmetry breaking carrying over to the QSL phase in the $S=1/2$ model~\cite{Kim-2008,Burnell-2009,Hickey-2017} sets the stage for a first realization in an insulator of a chiral spin liquid in three dimensions (see Refs.~\cite{Machida-2010,Lee-2013} for a metallic context). While we are unable to address this issue within the current implementation of PFFRG~\cite{supp}, an alternative route might be to proceed through a projective symmetry group classification of chiral spin liquids along with a variational Monte Carlo analysis~\cite{Bieri-2015,Bieri-2016}. 

While our DFT calculations show that {\lu} is well approximated by a Heisenberg Hamiltonian, it merely
serves as an effective minimal model. Indeed, a DM interaction term ${\sim}\mathbf{D}_{ij}{\cdot}(\mathbf{\hat{S}}_{i}{\times}\mathbf{\hat{S}}_{j})$~\cite{Dzyaloshinsky-1958} is also symmetry allowed. The Moriya rules~\cite{Moriya-1960} constrain this interaction to be one of two types, called ``direct'' or ``indirect''~\cite{Elhajal-2005,Kotov-2005}. Our DFT calculations of the DM term~\cite{Riedl-2016} find it to be ``indirect'' and estimate its magnitude to be ${\approx}0.08{-}0.1J_{1}$ (for a certain range of $U$ values). Within PFFRG, a treatment of the DM interaction for the pyrochlore lattice would be computationally expensive~\cite{Hering-2017}. However, a classical optimization calculation at $T=0$ shows that a $8\%-10\%$ DM interaction does not significantly alter the nature of the classical state of the pure Heisenberg model~\eqref{eqn:Hamil}. Indeed, we are unable to detect any shift in the Bragg peak positions within the available $\mathbf{k}$-space resolution, while only a minor redistribution of the spectral weight is observed. 

{\it Conclusion}. We have shown that {\lu} is well described by an ``extended'' Heisenberg model. Our PFFRG analysis shows that the system remains paramagnetic down to a temperature that is at least two orders of magnitude smaller compared to the Curie-Weiss temperature $\Theta_{\rm CW}$. The spin susceptibility profile displays momentum-dependent features forming a pattern of gearwheels. These signatures lend support to the view that the supposed quantum spin liquid could be viewed as a molten version of a parent classical magnetic order, which is found to be a noncoplanar incommensurate spiral. Our work provides a theoretical prediction for the characteristic spin susceptibility profile which should ultimately be compared with future neutron scattering experiments on single crystals. We hope that our work motivates further experimental investigations of the potentially very interesting Lu$_2$Mo$_2$O$_5$N$_2$ which may prove to be the first realization of a quantum spin liquid based on a spin-$\frac{1}{2}$ pyrochlore Heisenberg antiferromagnet, as our work here suggests by building on the report of Ref.~\cite{Clark-2014}.
 
We thank F. Becca, S. Bieri, L. Clark, I. I. Mazin, and J. Rau for useful discussions. The work was supported by the European Research Council through ERC-StG-TOPOLECTRICS-Thomale-336012. Y.I., T.M., and R.T. thank the DFG (Deutsche Forschungsgemeinschaft) for financial support through SFB 1170 (project B04). K.R. and R.V. thank the DFG for financial support through SFB/TR 49. J.R. is supported by the Freie Universit\"at Berlin within the Excellence Initiative of the German Research Foundation. S.R. is supported by the DFG through SFB 1143. The work at the University of Waterloo was supported by the Canada Research Chair program (M.G., Tier 1) and by the Perimeter Institute (PI) for Theoretical Physics. Research at the Perimeter Institute is supported by the Government of Canada through Innovation, Science and Economic Development Canada and by the Province of Ontario through the Ministry of Research, Innovation and Science. We gratefully acknowledge the Gauss Centre for Supercomputing e.V. for funding this project by providing computing time on the GCS Supercomputer SuperMUC at Leibniz Supercomputing Centre (LRZ).

%



\newcommand{\beginsupplement}{%
        \setcounter{table}{0}
        \renewcommand{\thetable}{S\arabic{table}}%
        \setcounter{figure}{0}
        \renewcommand{\thefigure}{S\arabic{figure}}%
        \setcounter{equation}{0}
        \renewcommand{\theequation}{S\arabic{equation}}%
        \setcounter{page}{1}
     }
 
 \bibliographystyle{apsrev4-1}
\renewcommand*{\citenumfont}[1]{S#1}
\renewcommand*{\bibnumfmt}[1]{[S#1]}
 
\newcommand\blankpage{%
    \null
    \thispagestyle{empty}%
    \addtocounter{page}{-1}%
    \newpage}

\blankpage

\chead{{\large \bf{---Supplemental Material---\\ Signatures of a gearwheel quantum spin liquid in a 
spin-$\frac{1}{2}$ pyrochlore molybdate Heisenberg antiferromagnet}}}

\thispagestyle{fancy}

\beginsupplement

\begin{figure}[b]
\includegraphics[width=\columnwidth]{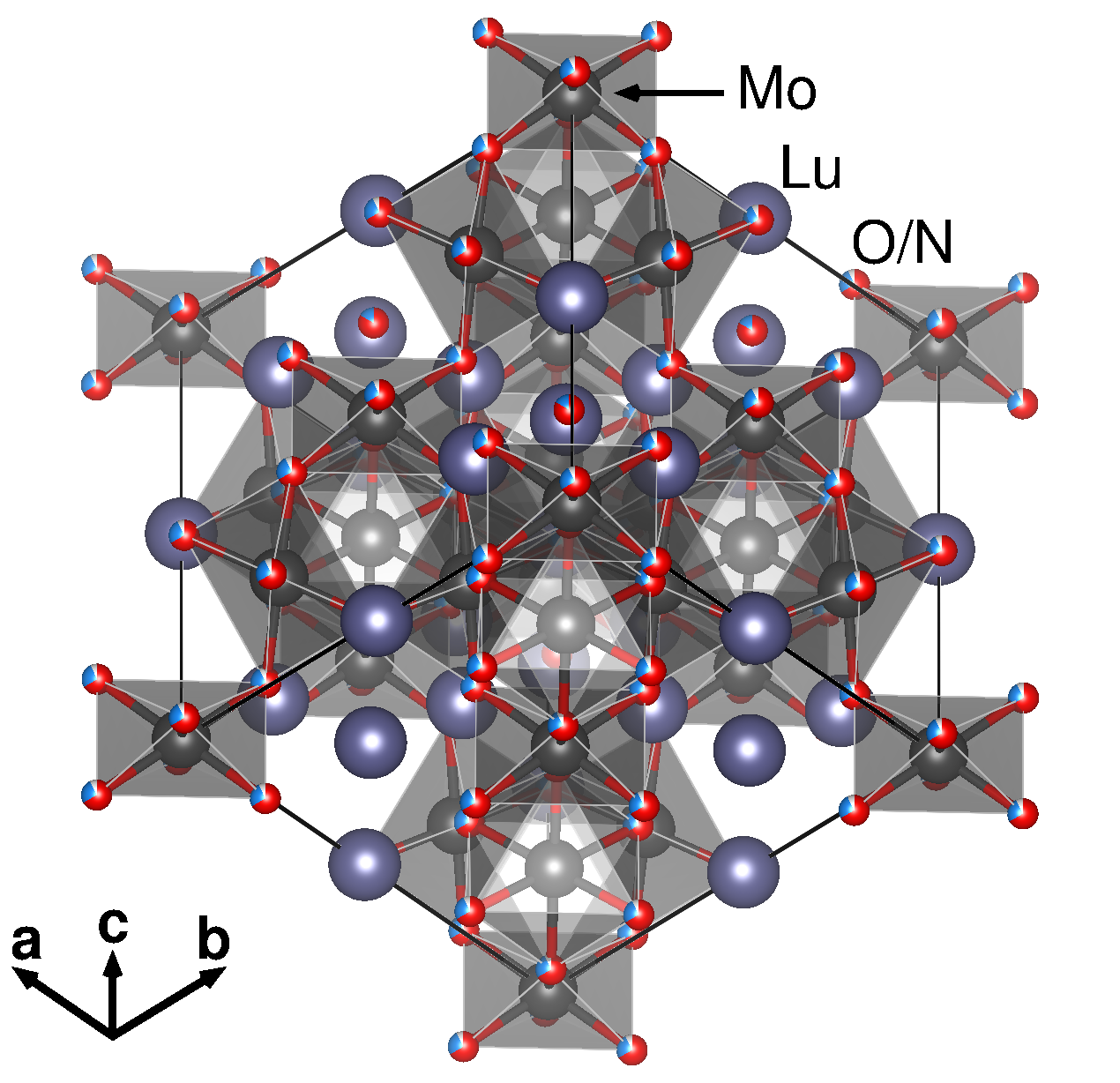}
\caption{Structure of {\lu}. Note that sites partially occupied by oxygen and nitrogen are shown by partly red, partly blue balls.}\label{fig:fullstructure}
\end{figure}

{\bf Structure---} We base our calculations on the {\lu} structure as determined by Clark
{\it et al.}~\cite{Clark-2014S} using powder neutron diffraction at
$T=4$~K. Both $48f$ and $8b$ positions of the pyrochlore structure are
partially occupied by oxygen and nitrogen [see Fig.~\ref{fig:fullstructure}]. Rietveld refinement yielded
O/N occupation numbers of 0.663/0.257 and 0.831/0.169 for the two
Wyckoff positions. While the $8b$ occupations add to one, the
refinement indicates slight O/N deficiency on $48f$. It is easily
determined that ideal occupations of $48f$ providing a 5:2 oxygen to
nitrogen ratio would be 0.6948/0.3052. In our calculations, we neglect
the possible O/N deficiency and adopt these ideal occupations of the
$48f$ position. Furthermore, we model the random O/N occupation of
$48f$ and $8b$ sites using the virtual crystal approximation~\cite{Bellaiche-2000S}. This
means that we assign nuclear charges of $Z=7.6948$ and $Z=7.831$ to
$48f$ and $8b$, respectively.

\begin{figure}[b]
\includegraphics[width=\columnwidth]{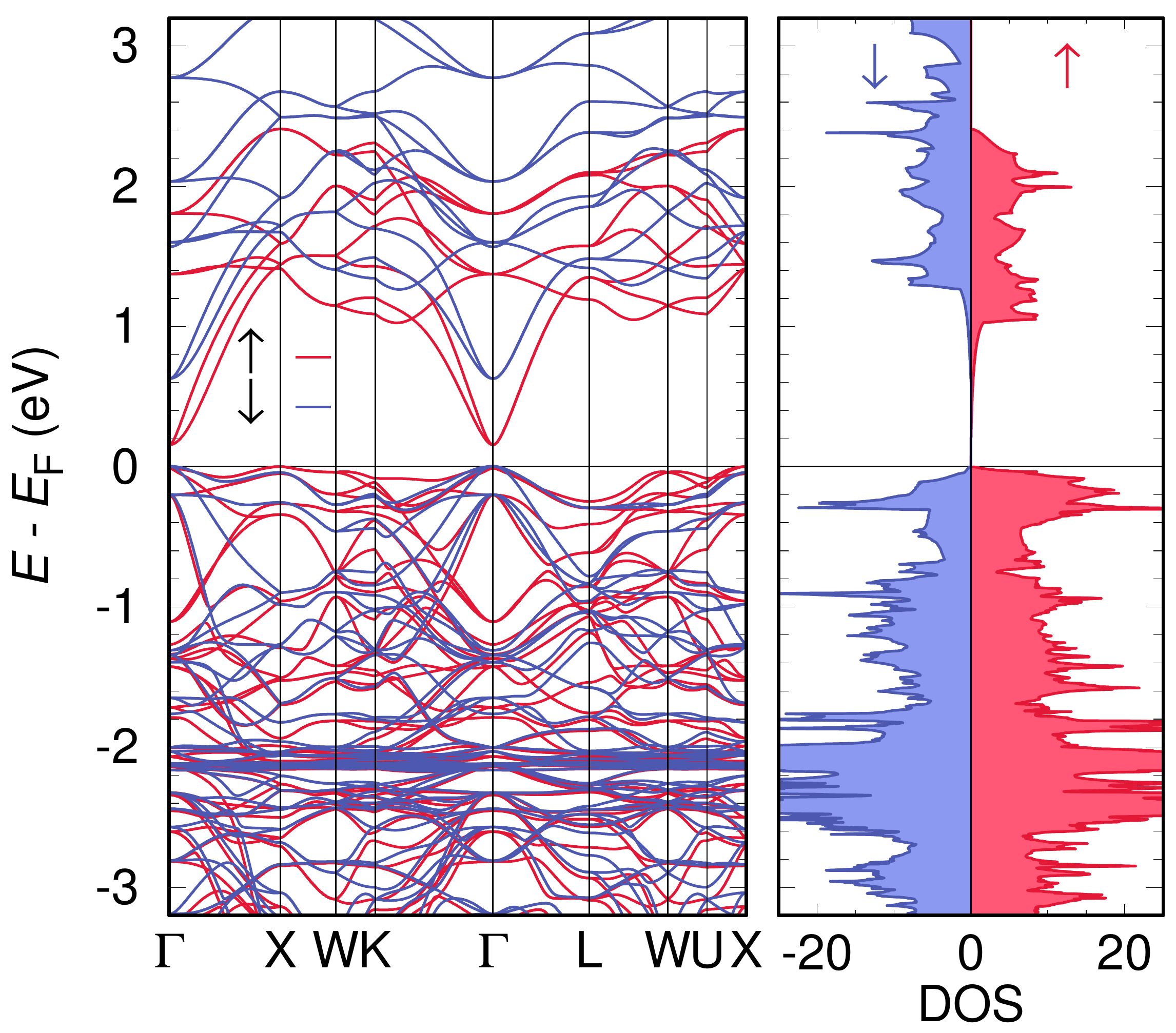}
\caption{Band structure and density of states of {\lu} calculated with
  GGA+$U$ functional at $U=2.5$~eV and $J_{\rm H}=0.6$~eV for the ferromagnetic state.  }\label{fig:bands}
\end{figure}

\begin{figure}[b]
\includegraphics[width=\columnwidth]{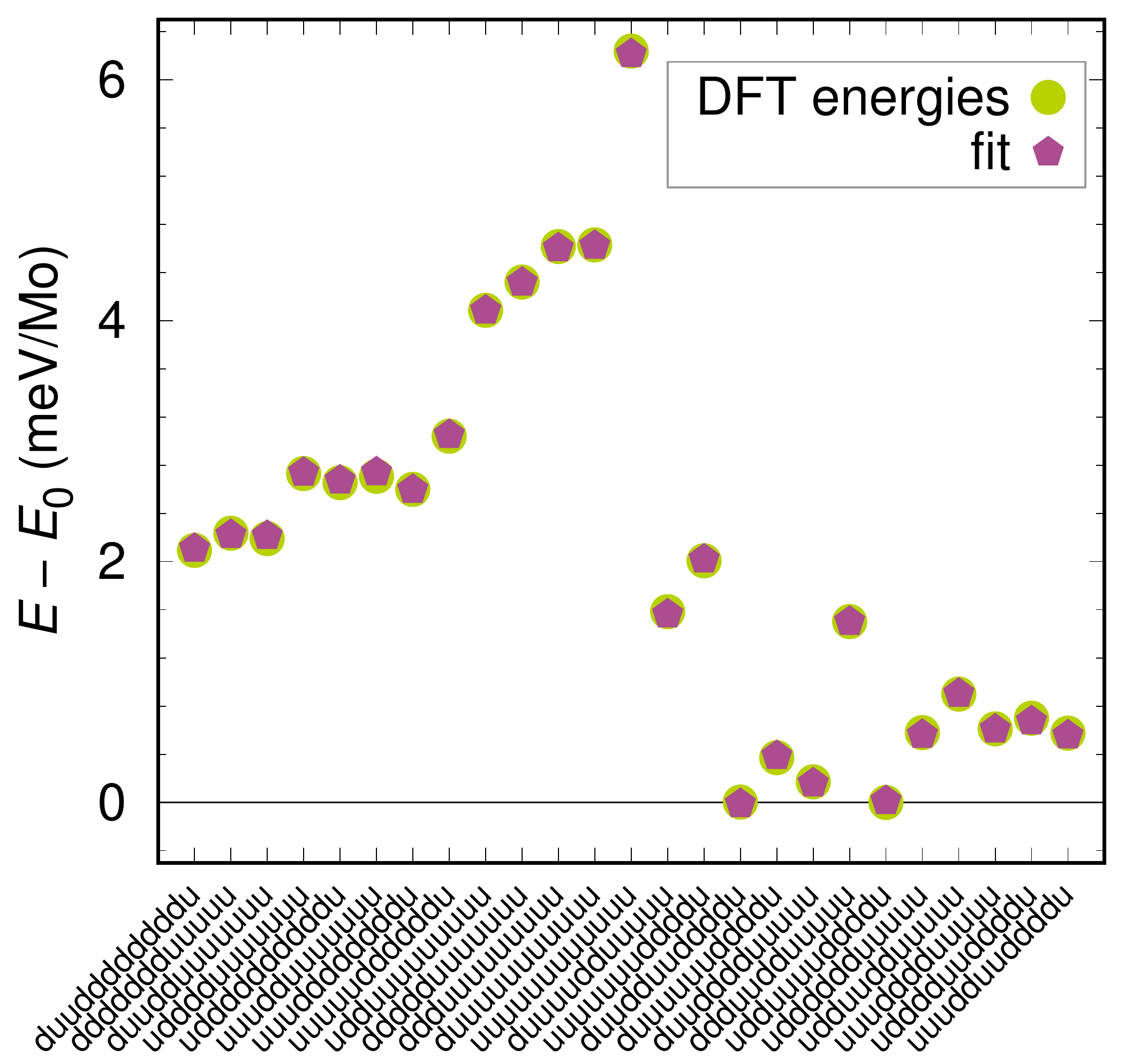}
\caption{Example for a set of 25 spin configurations of the considered
  $3\times 1\times 1$ supercell calculated with GGA+$U$ functional at
  $U=2.5$~eV and $J_{\rm H}=0.6$~eV. The quality of the fit to the
  Heisenberg model is very good.  }\label{fig:energies}
\end{figure}

\begin{table*}
\setlength{\tabcolsep}{12pt}
\centering
\begin{tabular}{lllllll}
 \hline \hline
       \multicolumn{1}{c}{$U$ (eV)}
    & \multicolumn{1}{c}{$J_{1}$ (K)}
    & \multicolumn{1}{c}{$J_{2}$ (K)}
    & \multicolumn{1}{c}{$J_{3{\rm a}}$ (K)}
    & \multicolumn{1}{c}{$J_{3\rm b}$ (K)}
    & \multicolumn{1}{c}{$J_{5}$ (K)}
    & \multicolumn{1}{c}{$\Theta_{\rm CW}$ (K)}  \\ \hline
       
\multirow{1}{*}{2} & 102.4(6) & $-0.1(5)$ & 23.2(5) & $-7.8(4)$ & $-1.4(2)$ & $-168(5)$ \\ 
\multirow{1}{*}{2.25} & 88.1(6) & 0.5(4) & 19.9(4) & $-6.6(4)$ & $-1.1(2)$ & $-147(5)$ \\ 
\multirow{1}{*}{$\mathbf{2.5}$} & $\mathbf{74.8(5)}$ & $\mathbf{0.6(4)}$ & $\mathbf{17.2(4)}$ & $\mathbf{-5.8(3)}$ & $\mathbf{-0.99(11)}$ & $\mathbf{-125(4)}$ \\ 
\multirow{1}{*}{2.75} & 62.0(5) & 0.6(3) & 15.0(4) & $-5.2(3)$ & $-0.89(10)$ & $-104(3)$ \\ 
\multirow{1}{*}{3} & 49.8(5) & 0.6(4) & 13.2(4) & $-4.8(3)$ & $-0.81(11)$ & $-84(4)$ \\ 
\multirow{1}{*}{3.25} & 37.8(5) & 0.6(4) & 11.7(4) & $-4.6(4)$ & $-0.74(11)$ & $-65(4)$ \\ 
\multirow{1}{*}{3.5} & 26.0(6) & 0.6(4) & 10.4(4) & $-4.4(4)$ & $-0.69(13)$ & $-46(4)$ \\ 
\multirow{1}{*}{3.75} & 14.2(6) & 0.5(5) & 9.3(5) & $-4.5(4)$ & $-0.64(14)$ & $-26(4)$ \\\hline \hline  

\end{tabular}
\caption{Exchange coupling constants for the oxynitride phase {\lu} determined from total energies of 25 spin configurations in a $3{\times}1{\times}1$ supercell using an $8{\times}8{\times}8$ $\mathbf{k}$-mesh [see Fig.~\ref{fig:structure}]. The parameters corresponding to $U=2.5$ eV (marked in bold) are used for the PFFRG simulations. We adopt the convention in which each pair $\langle i,j\rangle$ in the summation in the exchange Hamiltonian [Eq.~\ref{eqn:Hamil}] is counted only once. Accordingly, the formula for the Curie-Weiss temperature is $\Theta_{\rm CW}=-\frac{1}{3}S(S+1)\sum_{n} {\rm z}_{n} J_{n}$, where the summation extends over all neighbors with which a given spin interacts, and ${\rm z}_{n}$ is the coordination number at the $n$th-nearest-neighbor~\cite{Khomskii-2010S}.}
\label{tab:DFT}
\end{table*}

{\bf Electronic structure---} We perform electronic structure calculations for {\lu} using the full
potential local orbital (FPLO) code~\cite{Koepernik-1999S} using the generalized
gradient approximation (GGA) functional in its Perdew-Burke-Ernzerhof (PBE) form~\cite{Perdew-1996S}.  We
correct for the strong correlations on the Mo$^{5+}$ $4d$ orbitals
using the GGA+$U$ method~\cite{Liechtenstein-1995S}.  Fig.~\ref{fig:bands} shows the
electronic structure for a ferromagnetic solution calculated with
GGA+$U$. The Hund's rule coupling is fixed at a value of $J_{\rm
  H}=0.6$~eV, which is typical for $4d$ transition metal ions. The
onsite interaction is chosen to be $U=2.5$~eV because the Heisenberg
Hamiltonian parameters estimated at this value yield a Curie-Weiss
temperature which is close to the experimentally observed value
$\Theta_{\rm CW}=-121(1)$~K. There are many bands as the primitive cell
contains two formula units of {\lu}. Per Mo$^{5+}$ ion, there is one
occupied band of $4d$ character in the majority channel ($\uparrow$),
corresponding to a magnetic moment of precisely $S=1/2$. The narrow
bands around $-2$ eV are the occupied Lu $4f$ states. The other occupied
bands are O/N $2p$. At this value of $U$, {\lu} is a semiconductor
with a small gap of $E_g=0.15$~eV.  

{\bf Exchange couplings---} Next, we calculate the total
energies for 25 different spin configurations of a $3\times 1\times 1$
supercell of the primitive cell of {\lu}. An example for this
procedure is illustrated in Fig.~\ref{fig:energies}. We obtain the
estimates for the Heisenberg exchange parameters listed in
Table~\ref{tab:DFT} by fitting the DFT+$U$ total energies against the
classical energies of the Heisenberg Hamiltonian.  The evolution of
exchange couplings with onsite interaction $U$ is shown in
Fig.~\ref{fig:couplings}. While the next nearest neighbor coupling
$J_2$ is negligibly small, the two inequivalent third neighbor
couplings $J_{3\rm a}$ (connecting two Mo$^{5+}$ sites with with a nearest
neighbor in between) and $J_{3\rm b}$ (across an empty hexagon in one of
the three interpenetrating kagome lattices of the pyrochlore
structure) are substantial and of different sign; $J_{3\rm a}$ is
antiferromagnetic like $J_1$, and $J_{3\rm b}$ is ferromagnetic. We do not
expect exchange couplings at Mo-Mo distances of 8~{\AA} or more to
play a major role. The $3\times 1\times 1$ supercell does not allow us
to resolve $J_4$ ($d_{\rm Mo-Mo}=8.02$~{\AA}) but we can determine
$J_5$ ($d_{\rm Mo-Mo}=9.49$~{\AA}) and find it to be very small. We derived the anisotropic exchange couplings in the framework of a combination of relativistic DFT calculations with exact diagonalization of a generalized Hubbard Hamiltonian on finite clusters, detailed in Ref.~\cite{Riedl-2016S}. Note that $U$ in this method does not enter in
the same way as in the GGA+$U$ total energy calculations. We obtain the estimate of $|D|/J$ by scanning $U$ values of up to 3.6 eV and values of the Hund's rule coupling $J_{\rm H}$ in the range of 0.6 eV to 0.8 eV.

{\bf Pseudofermion FRG---}
The PFFRG scheme~\cite{Reuther-2010S,Reuther-2011aS,Reuther-2011bS,Reuther-2011cS,Reuther-2011dS,Reuther-2014S} is a non-perturbative framework capable of handling arbitrary two-body spin-interactions of both diagonal and off-diagonal type~\cite{Iqbal-2016cS,Hering-2017S}, with any given spin~\cite{Baez-2016S}. It is formulated in the SU(2) fermionic representation of spins, which amounts to rewriting the physical spin operator at each site in terms of Abrikosov pseudofermions, 
\begin{equation}
\mathbf{\hat{S}}_{i}=\frac{1}{2}\sum_{\alpha,\beta}\hat{f}^{\dagger}_{i,\alpha}\pmb{\sigma}_{\alpha\beta}\hat{f}_{i,\beta}\;,
\end{equation}
where $\alpha$, $\beta=\uparrow$ or $\downarrow$, and $\hat{f}^{\dagger}_{i,\alpha}$ ($\hat{f}_{i,\alpha}$) are the pseudofermion creation (annihilation) operators, and $\pmb{\sigma}$ is the vector of Pauli matrices. The fermionic representation is endowed with an enlarged Hilbert space which includes the unphysical {\it empty} and {\it doubly}-occupied sites carrying zero-spin, and must be projected out to restore the original Hilbert space of the Heisenberg model which has one-fermion-per-site. One way to achieve this is to add onsite level repulsion terms $-A\sum_i {\bf S}_i^2$ to the Hamiltonian, where $A$ is a positive constant~\cite{Baez-2016S}. Such terms lower the energy of the physical states but do not effect the unphysical ones. As a consequence, at sufficiently large $A$ the low energy degrees of freedom of $H$ are entirely within the physical sector of the Hilbert space. For a wide class of spin systems (including the models considered here) one finds that even for $A=0$, the ground state of the fermionic Hamiltonian obeys the one-fermion-per-site constraint~\cite{Baez-2016S}. This is because unphysical occupations effectively act like a vacancy in the spin lattice, and are associated with a finite excitation energy of the order of the exchange couplings. As a consequence, the ground state of the fermionic system is identical to the ground state of the original spin model where each site is singly occupied.

Within PFFRG, a step-like infrared frequency cutoff $\Lambda$ along the Matsubara frequency axis is introduced in the bare fermion propagator $G_0(i\omega)=\frac{1}{i\omega}$, i.e., $G_0(i\omega)$ is replaced by
\begin{equation}
G_0^\Lambda(i\omega)=\frac{\Theta(|\omega|-\Lambda)}{i\omega}\;.
\end{equation}
Implanting this modification into the generating functional of the one-particle irreducible vertex function and taking the derivative with respect to $\Lambda$ yields an exact but infinite hierarchy of coupled flow equations for the $m$-particle vertex functions~\cite{Metzner-2012S}, which constitutes the FRG ansatz. The first two equations for the self energy $\Sigma^\Lambda$ and the two-particle vertex $\Gamma^\Lambda$ have the forms
\begin{equation}\label{eq:FirstFlow}
 \frac{d}{d\Lambda}\Sigma^{\Lambda}\left(1';1\right)=-\frac{1}{2\pi}\sum\limits_{2'\,2}\Gamma^{\Lambda}\left( 1',2';1,2\right)S^{\Lambda}\left(2,2'\right)
 \end{equation}
 and
 \begin{align}
  \frac{d}{d\Lambda}\Gamma^{\Lambda}\left(1',2';1,2\right)&=\frac{1}{2\pi}  \sum\limits_{3'\,3} \Gamma_3^\Lambda \left( 1',2',3';1,2,3\right) S^{\Lambda}\left(3,3'\right) \nonumber \\
 +\frac{1}{2\pi}\sum\limits_{3'\,3\,4'\,4}\:&\Bigg[ \Gamma^{\Lambda}\left( 1',2';3,4\right) \Gamma^{\Lambda}\left( 3',4';1,2\right) \nonumber \\ -\Gamma^{\Lambda}( 1',4';1&,3) \Gamma^{\Lambda}\left( 3',2';4,2\right) - \left (3'\leftrightarrow 4', 3 \leftrightarrow 4 \right ) \nonumber \\ +\Gamma^{\Lambda}( 2',4';1&,3) \Gamma^{\Lambda}\left( 3',1';4,2\right) + \left (3'\leftrightarrow 4', 3 \leftrightarrow 4 \right ) \Bigg ] \nonumber \\ \label{eq:SecondFlow} \times G^{\Lambda}(3,3')&S^{\Lambda}(4,4')\;,
\end{align}
where $\Gamma_3^\Lambda$ denotes the three-particle vertex. Here, $G^\Lambda$ is the fully dressed propagator and $S^\Lambda$ is the so-called single scale propagator defined by
\begin{equation}
\label{eq:SingleScale}
 S^{\Lambda}=G^{\Lambda} \frac{d}{d\Lambda}\left[ G^{\Lambda}_0 \right ]^{-1} G^{\Lambda}\;.
\end{equation}
Note that the arguments $1,2,\ldots$ of the vertex functions and propagators denote multi indices ``$1\equiv\{\omega_1,i_1,\alpha_1\}$'' containing the frequency variable $\omega_1$, the site index $i_1$ and the spin index $\alpha_1$.

For a numerical solution, this hierarchy of equations is truncated to keep only the self-energy $\Sigma^\Lambda$ and two-particle vertex $\Gamma^\Lambda$. Particularly, the truncation on $\Gamma_3^\Lambda$ is performed such that, via self-constistent feedback of the self-energy into the two-particle vertex, the approach remains separately exact in the large $S$ limit as well as in the large $N$ limit [where the spins' symmetry group is promoted from SU$(2)$ to SU$(N)$]~\cite{Baez-2016S}. While the terms representing the large $S$ limit [second line of Eq.~(\ref{eq:SecondFlow})] describe the long-range ordering in classical magnetic phases, the large $N$ terms [fourth line of Eq.~(\ref{eq:SecondFlow})] characterize the system with respect to non-magnetic resonating valence bond or dimer crystal phases. This allows for an unbiased investigation of the competition between magnetic ordering tendencies and quantum paramagnetic behavior. Approximations due to the neglect of the three-particle vertex $\Gamma_3^\Lambda$ concern subleading orders in $1/S$ and $1/N$. Such terms are essential for probing possible chiral correlations in paramagnetic phases, e.g., in chiral spin liquids with a scalar chiral order parameters of the form $\sim\langle ({\bf S}_i\times{\bf S}_j)\cdot{\bf S}_k\rangle$. Therefore, the current implementation of the PFFRG does not allow to describe the possibility of a spin system to form chiral spin liquids.

The two-particle vertex in real space is related to the static spin-spin correlator
\begin{equation}
\chi_{ij}^{\mu\nu}=\int_0^\infty d\tau\langle\hat{S}_i^\mu(\tau)\hat{S}_j^\nu(0)\rangle\label{correlator}
\end{equation}
where $\hat{S}_i^\mu(\tau)=e^{\tau\hat{\mathcal{H}}}\hat{S}_i^\mu e^{-\tau\hat{\mathcal{H}}}$. As a finite-size approximation, correlators $\chi^{\mu\nu}_{ij}$ are only calculated up to a maximal separation between sites $i$ and $j$. The main physical outcome of the PFFRG are the Fourier-transformed correlators, i.e., the static susceptibility $\chi^{\mu\nu,\Lambda}({\bf k})$ evaluated as a function of the RG scale $\Lambda$, which in three dimensions (for a $S=1/2$ system) is related to a temperature $T=(\frac{\pi}{2})\Lambda$~\cite{Iqbal-2016bS}. In our case, the maximal distance of the correlators is $\sim11.5$ lattice spacings corresponding to a total volume of $2315$ correlated sites which ensures a proper {\bf k}-space resolution. We implement an approach in which despite spatially limited vertices the system size is assumed to be, in principle, infinitely large. The frequency dependence of the two-particle vertex function is discretized over 64 points. If a system develops magnetic order, the corresponding two-particle vertex channel anomalously grows upon decreasing $\Lambda$ and eventually causes the flow to become unstable. Otherwise, a smooth flow behavior of the susceptibility down to $\Lambda\rightarrow0$ signals the absence of magnetic order.

{\bf Iterative minimization of the classical Hamiltonian---}
The ground state of a classical Heisenberg Hamiltonian is found using an iterative minimization scheme which preserves the fixed spin length constraint at every site~\cite{Lapa-2012S}. In contrast, within the Luttinger-Tisza method the fixed spin length  constraint is only enforced globally, i.e., $\sum_{i}|\mathbf{S}_{i}^{2}|=S^{2}N$, where $N$ is the total number of lattice sites, implying that local moment fluctuations which are now permissible take us beyond the classical approximation by approximately incorporating some aspects of the quantum Hamiltonian~\cite{Kimchi-2014S}. Starting from a random spin configuration on a lattice with periodic boundary conditions, we choose a random lattice point and rotate its spin to point antiparallel to its local field defined by
\begin{equation}
 \mathbf{h}_i = \frac{\partial H}{\partial \mathbf{S}_i} = \sum\limits_{j\neq i} J_{ij} \mathbf{S}_j.
\end{equation}
This results in the energy being minimized for every spin update and thereby converging to a local minimum. We choose a lattice with $L=32$ cubic unit cells in each direction, and thus a single iteration consists of $16 L^3$ sequential single spin updates. This iterative scheme is repeated many times starting from different random initial configurations to maximize the likelihood of having found a global minimum. From the minimal energy spin configuration, the spin structure factor 
\begin{equation}
\mathcal{F}(\mathbf{k}) = \frac{1}{16 L^3} \left|\sum\limits_i \mathbf{S}_i e^{\imath \mathbf{k} \cdot \mathbf{r}_{i}}\right|^{2}
\end{equation}
is computed.

\end{document}